\documentclass[aps,prl,twocolumn,showpacs,preprintnumbers,amsmath,amssymb,superscriptaddress]{revtex4}
\usepackage{graphicx}% Include figure files
\usepackage{dcolumn}% Align table columns on decimal point
\usepackage{color}

\usepackage{amsmath}
\usepackage{amssymb}

\def\bea{\begin{eqnarray}}
\def\eea{\end{eqnarray}}
\def\ben{\begin{equation}}
\def\een{\end{equation}}
\def\benu{\begin{enumerate}}
\def\enu{\end{enumerate}}

\def\sss{\scriptscriptstyle\rm}

\def\1var{(\bx_1...\bx\N)}

\def\br{{\bf r}}

\def\bx{{x}}

\def\s{_{\sss S}}
\def\xc{_{\sss XC}}

\def\Hxc{_{{\sss HXC}}}

\def\N{_{\sss N}}

\def\sph_int{ {\int d^3 r}}

\begin{document}
%\large\sf

%\preprint{UCI DFT GROUP: preprint FWZB07}

\title{Time-dependent Density Functional calculation of e-H scattering}
\author{Meta van Faassen}
\affiliation{Department of Chemistry and Chemical Biology, Rutgers
University, 610 Taylor Road, Piscataway, NJ 08854-8087, USA}
\author{Adam Wasserman} 
\affiliation{Department of Chemistry and Chemical Biology, Harvard University, 12 Oxford St., Cambridge Massachusetts 02138, USA}
\author{Eberhard Engel}
\affiliation{Center for Scientific Computing, JW Goethe-Universit{\"a}t Frankfurt, Max-von-Laue-Stra\ss{}e 1, D-60438 Frankfurt/Main, Germany}
\author{Fan Zhang}
% Kieron: affiliation changed
\affiliation{Department of Chemistry and Chemical Biology, Rutgers
University, 610 Taylor Road, Piscataway, NJ 08854-8087, USA}
\author{Kieron Burke}
\affiliation{Department of Chemistry, University of California, Irvine, CA 92697, USA}

\date{\today}
\begin{abstract}
Phase shifts for single-channel elastic electron-atom scattering are derived from time-dependent density functional theory. The H$^-$ ion is placed in a spherical box,
its discrete spectrum found, and phase shifts deduced. Exact-exchange yields an excellent approximation to the ground-state Kohn-Sham potential, while the adiabatic local density approximation yields good singlet and triplet phase shifts.
\end{abstract}


\maketitle

Modern density functional theory (DFT) ~\cite{HK64,KS65,FNM06} has proven very succesful in quantum
chemistry and solid-state physics. The time-dependent
formulation, TDDFT~\cite{RG84}, extends this
success to excited-state
properties~\cite{BWG05}.
Thus, excitation energies and oscillator strengths
of electronic transitions of atoms, molecules, and clusters are now routinely calculated via TDDFT within, e.g., the adiabatic
local density approximation (ALDA)~\cite{MUNR06}. 
But such calculations
are almost exclusively for optical response to either weak~\cite{BWG05}
or strong~\cite{MG04} fields.

The problem of calculating low-energy elastic electron scattering from atoms and molecules
is demanding, and solving the Schr\"odinger equation for continuum states of polyatomic
molecules can be expensive.  However, such solutions are needed for the
emergent
field of electron-impact chemistry~\cite{HGDS03}, especially since recent experiments show that
low energy electrons can cleave DNA~\cite{BCHH00,CS03}.  Through efficient use of R-matrix theory, calculations
within static exchange (amounting to scattering from an effective one-body potential) have
been performed for a single DNA base~\cite{TG06}.   A TDDFT approach could prove highly
useful here, allowing the incorporation of correlation effects with little additional
cost beyond the original scattering calculation.

With this ultimate goal in
mind, we demonstrate a simple method for using TDDFT to calculate phase shifts.
We find the continuum states of the $N+1$-electron
problem, where the target has $N$ electrons.
Our method is extremely practical in spherical cases, such as atoms.
It is based on a little-used formula~\cite{F35,FL73,L74,S74,S06}
(exact for finite-ranged potentials)
that relates the phase shift of the continuum problem to discrete energies 
of the same potential, but placed inside
a box  whose edge is beyond the range of the potential.
This formula bypasses many of the complications
of our original work~\cite{WMB05}, as now we need only
find bound-bound transition energies,
where TDDFT has already proven
successful.  Furthermore, since our general approach requires that 
the $N+1$-electron system be bound, by putting the system in a box,
our new method can be applied, at least in principle, even when the `ground state' of the
$N+1$-electron system is
only a resonance.

A vital element in any DFT approach is the accuracy
of approximate functionals used.
In this sense, electron scattering from the H atom
is a very severe test, since H$^-$ (the $N+1$-electron system)
is so strongly correlated.
The underlying ground-state Kohn-Sham (KS) potential
is crucial to any TDDFT calculation, especially for atoms, and is
known essentially exactly for H$^-$
~\cite{UG93}.  We find
that exact-exchange, as calculated in an optimized effective potential
(OEP) code,
yields very accurate KS phase-shifts, i.e., very close to
those of the known exact KS potential.  Next, we show that the ALDA, the workhorse of TDDFT, yields very
good shifts for both singlet and triplet (TD-spin-DFT) scattering.
Thus, we demonstrate that a simple formalism allows scattering calculations from TDDFT; that modern approximations yield sufficiently accurate ground-state potentials; and that standard TDDFT approximations are sufficiently accurate. We perform the first such calculation on the prototype target, the H atom.

We begin with some exact observations about scattering from
a potential.  Consider a spherical potential that
has a finite range, i.e., $v(r)=0$ beyond some radius $R_c$.
Now imagine inserting a hard wall at any $R_b > R_c$, not necessarily
far beyond $R_c$, and solving for the bound states.
Any such solution is in fact a solution to the original scattering
problem that happens to have a node right at $R_b$.
Study of the wavefunction between $R_c$ and $R_b$ to identify the
phase shift yields:
\ben
\tan(\delta_{l\alpha}) = -j_l(k_\alpha\,R_b)/\eta_l (k_\alpha\,R_b)
\label{e:phase}
\een
where $j_l$ and $\eta_l$ are the two free-space solutions to the
radial Schr\"odinger equation, i.e., the spherical Bessel and von Neumann
functions, 
$k_\alpha = {\sqrt{2 E_\alpha}}$, and $E_\alpha$ is
the $\alpha$-th eigenenergy.
For s-wave scattering, Eq.~(\ref{e:phase}) reduces to:
\ben\label{e:phase2}
\delta_\alpha = - k_\alpha\ R_b + \alpha \pi \quad (l=0)
\een
The phase shifts are only determined modulo $\pi$, but we have
added $\alpha\pi$, the free-particle value of $k_\alpha\,R_b$, so
that all shifts are relative to 0. For any given $l$ and $R_b$, this method yields the phase shift at
a discrete set of energies.  
%
% Referee A is confused about the following sentence:
% "By increasing $R_b$ by a small amount, sufficient
% to add one more positive energy state, {\em all} values of $k$ can be
% obtained, {\em except} the very lowest." 
%
%   Referee A: confused about lowest value extractable
%
% Explain for Referee A why the graphs go to zero
%
For a fixed potential, starting from {\em any} $R_b$ value, $R_0$, one can continuously increase $R_{b}$ to about $2R_0$ and generate $\delta$ at {\em all} energies above a minimum $E_{\rm min}=\frac{1}{2}[(\pi-\delta_n)/R_b]^2$, where $R_b$ is the largest box used.
Usually $R$-matrix theory is more
 convenient, as it does not require the wavefunction to have a node at the box radius,
 and  so all energies can be found with just one value of $R_{b}$. But it relies on knowing the logarithmic derivative of the wavefunction at $R_{b}$, which is not available in TDDFT.

\begin{figure}[tbp]
\includegraphics{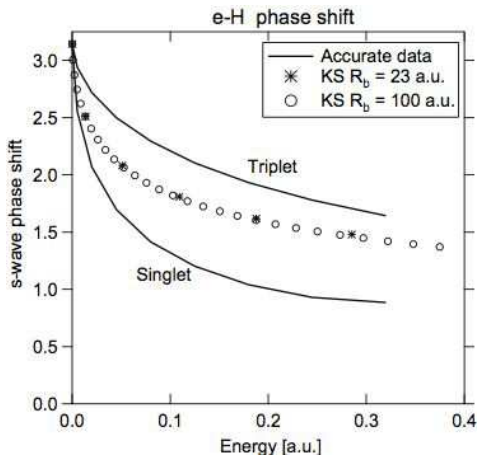}
% Referee B: change 15 a.u. in 23 a.u. (as in figure legend)
\caption{\label{f:KSphase}Accurate quantum chemical singlet and triplet s-phase shifts~\cite{S61}, together with the KS values, calculated with a wall at %15 a.u.
23 a.u. and at 100 a.u..}
\end{figure}
To illustrate the method, and show how useful the exact ground state KS potential is, in Fig. ~\ref{f:KSphase} we plot accurate quantum calculations for both singlet and
triplet elastic scattering from  hydrogen~\cite{S61}.   We also plot the result of potential
scattering from the exact ground state KS potential of H$^-$.  This was found by Umrigar {\em et al.}~\cite{UG93}, from an extremely accurate quantum Monte-Carlo calculation for the ground state of H$^-$, calculating the density, and finding $\upsilon_s(\br)$ by inverting the KS equation. We obtained the positive orbital energies (necessary to evaluate Eq.~(\ref{e:phase2}))  from a well-established fully numerical
spherical DFT code, which includes the optimized effective potential method (OEP) and has been supplemented by the option to insert a hard-wall at a distance
$R_b$ from the origin\cite{JE05}.

The KS phase shift fits between the two curves, just as the pure KS orbital energy differences lie between singlet and triplet excitations for He~\cite{FB06,F06}.  The calculations at two
values of $R_b$ demonstrate the results are independent of $R_b$.  We choose the wall far from the origin to ensure the self-consistent
{\em ground-state} results are not affected by its position and to approach zero energy, but we emphasize the fact that Eq.~(\ref{e:phase2}) is exact for any $R_b>R_c$.

\begin{figure}[tpb]
\includegraphics{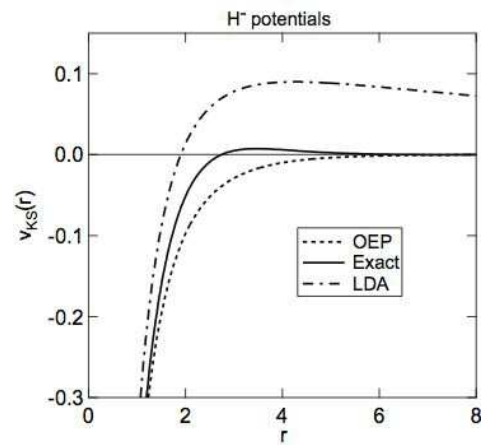}
\caption{\label{f:pots}The exact, exact-exchange, and LDA KS-potentials for H$^-$.}
\end{figure}
\begin{figure}[tpb]
\includegraphics{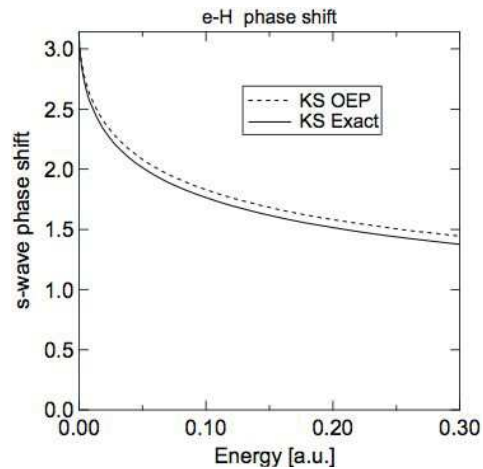}
\caption{\label{f:scatt}The s-wave phase shifts for the exact and exact-exchange KS potentials for H$^-$.}
\end{figure}
Obviously, the decay of the ground-state KS potential is crucial to the accuracy of
this method, and in any practical application, the exact KS potential is unavailable. Therefore we study the behavior of two approximate potentials, exact-exchange (OEP) and the local density approximation (LDA).  In Fig.~\ref{f:pots}, we plot both the exact and approximate KS potentials for H$^-$.   The LDA potential is far too shallow, a well-known failing of most commonly-used approximations to ground-state DFT. The true self-consistent LDA potential does not support any bound states, so to obtain the potential we put the system in a large box, forcing the states to be bound~\cite{SRZ77}.
Thus, the LDA potential is utterly unsuitable for this type of calculation. On the other hand,
the exact-exchange potential decays correctly as $r\to\infty$, missing only the small positive correlation potential for small $r$. 
Many modern $R$-matrix based methods
start with the nuclear potential and
the pure electrostatic (i.e., Hartree) potential, and then add the LDA exchange potential from DFT, i.e., the Slater contribution that decays exponentially, as $n^{1/3}(r)$.
Since this potential misses the correct asymptotic behavior, $v\xc(r) \to -1/r$, a `polarization' potential must be added~\cite{M83}.
Our KS potentials, either exact or exact-exchange, already have the correct asymptotic behavior, i.e., they contain
the polarization potential.  Without this feature, our KS potentials
would have the wrong asymptotic behavior, and would not be long-ranged
for neutral atoms. In Fig.~\ref{f:scatt}, we plot the scattering from the exact and exact-exchange potentials, demonstrating that exact-exchange, as is now available in many
codes~\cite{DG01}, is perfectly adequate for this purpose.

%%%%%%%%
% Referee A: add a sentence to make transition smoother
% Referee B: needs more explanation about singlet/triplet
% changed to spin decomposed formalism
To go even further, e.g. to account for singlet-triplet splitting, we must use TDDFT. Within the  formalism of TDDFT within linear response we can, in principle, obtain the true singlet and triplet excitation energies, and thus the phase shifts.
  We label all single-particle
excitations from the ground to unoccupied excited states via $q=(i,a)$, where
$i$ implies occupied, $a$ implies unoccupied, and define $\Phi_{q\sigma}(\br)  =
\phi_{i\sigma}^*(\br)\phi_{a\sigma}(\br)$, where $\sigma$ is a spin index and $\phi_i(\br)$ is an eigenstate of the ground state $v\s(\br)$. 
Casida~\cite{C96} cast
the TDDFT response equations as a eigenvalue equation
\begin{equation}
\sum_{q'} {\tilde\Omega}_{q\sigma q'\sigma'} (\omega)\ a_{q'\sigma} = \omega^2\ a_{q\sigma},
\label{Casida}
\end{equation}
where
\bea
{\tilde\Omega}_{q\sigma q'\sigma'}(\omega)
&=& \omega^2_{q\sigma} \delta_{qq'}\delta_{\sigma\sigma'} \nonumber\\
&+& 2 {\sqrt{\omega_{q\sigma}\omega_{q'\sigma'}}}
\langle q\sigma | f^{\sigma\sigma'}\Hxc (\omega) | q'\sigma' \rangle.
\label{Odef}
\eea
and $\langle q\sigma | f^{\sigma\sigma'}\Hxc (\omega) | q'\sigma' \rangle$ is the matrix element
of the Hartree-XC kernel in the set of functions $\Phi_{q\sigma}(\br)$. We also defined 
$\omega_{q\sigma}=\epsilon_{i\sigma}-\epsilon_{a\sigma}$, where $\epsilon_{i\sigma}$ is the KS orbital energy of state $i$ with spin $\sigma$. 
The XC kernel is the functional derivative of the XC potential in TDDFT~\cite{MG04,BWG05}
%Referee B: make it clear the ALDA is used
and we assume the frequency independent ALDA kernel in the following.
%
% For Referee B, again stress that in practice we obtain both singlet and triplet energies
% reference to Casida paper
The ${\tilde\Omega}$-matrix can be split in separate singlet and triplet ${\tilde\Omega}$-matrices~\cite{C96}. 
Solving Eq.~(\ref{Casida}) therefore yields predictions of both singlet and triplet transition
frequencies, $\omega$.
In order to perform these calculations we have added subroutines to evaluate the
matrix elements needed for a TDDFT calculation. Since the system studied is small, we exactly diagonalize the ${\tilde\Omega}$-matrix.
%%%%%%%%%%

\begin{figure}[tpb]
\includegraphics{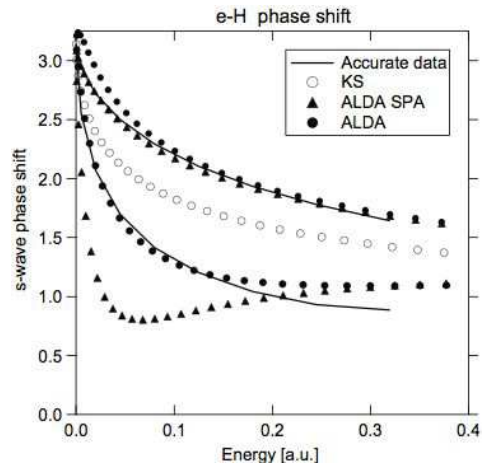}
\caption{\label{f:TDDFTphase}Singlet and triplet TDDFT curves from an SPA and full ALDA calculation, together with the KS values and accurate quantum chemical data from Ref.~\cite{S61}. The ground state KS potential is exact-exchange. The wall location in all calculations is at 100 a.u.}
\end{figure}
In Fig.~\ref{f:TDDFTphase}, we show the results obtained from a TDDFT ALDA calculation,
% Referee A: mention that the ground state of the TDDFT is always OEP in our case
 but using the OEP ground-state potential.
Apart from the full results we also show results obtained with the single-pole approximation (SPA), which ignores the off-diagonal matrix elements in Eq.~(\ref{Odef})~\cite{PGG96}. The SPA is analogous to the distorted-wave Born approximation used in earlier work, which worked well for electron scattering from He$^+$~\cite{WMB05}, but fails
badly for H. In Fig.~\ref{f:TDDFTphase} the SPA is indeed a poor approximation to the full singlet curve, especially at low energies.
For the triplets, on the other hand, we obtain excellent results that are on top of the reference values. 

If we now look at the full calculation we see that including the off-diagonal matrix elements considerably improves the singlet values giving results very close to the reference data. For the triplet the results only change for smaller energies  where the values are too big and there is a small  ``bump'' close to $E=0$. 
We believe this bump to be unphysical, due to coupling among transitions being
treated incorrectly by our approximate XC-kernel, as $E\to0$. This suspicion is reinforced by the fact that in this region, the full ALDA triplet results depend on the position of the wall, and so cannot be trusted. 
% For Referee A: add a sentence
However, similar effects were found with other common kernels, such as exact exchange, so we believe some delicate behavior of the XC kernel is required to avoid this artifact.

\begin{table}[tpb]
\caption{TDDFT s-wave scattering lengths.}
\begin{ruledtabular}
\begin{tabular}{ccc}
 & Singlet $a$ & Triplet $a$ \\
Accurate data\footnotemark[1] &  5.97  & 1.77  \\
ALDA SPA    &  9.7   &  1.8   \\
ALDA &  5.6   &  2.0\footnotemark[2]
\end{tabular}
\end{ruledtabular}
\footnotetext[1]{Accurate variational calculations from~\cite{S61}}
\footnotetext[2]{This is the value as obtained from our tangent approximation as explained in the text}
\label{t:scattlength}
\end{table}%
To quantify results for low energies, the scattering length is defined by the effective range expansion,
\begin{equation}
k^{2l+1}\cot\delta_{l}(k){\mathop=\limits_{k \to 0}} -\frac{1}{a_{l}}+\frac{1}{2}r_{el}k^{2}+O(k^{4}),
\end{equation}
where $a_{l}$ is the scattering length and $r_{el}$ the effective range. Since we have no wave function, to extract $a_l$ we must fit our data to the above expression to obtain the scattering lengths. We give a rough estimate of the expected TDDFT scattering lengths in Table~\ref{t:scattlength}, by fitting our results for small $k$. As a reference, the KS scattering lengths are 4.7 for the exact potential and 4.2 for the exact-exchange potential.
%\begin{figure}[tpb]
%\includegraphics[width=8.5cm]{Tfit}
%\caption{\label{f:Tfit}The solid circles are triplet ALDA values for a wall at 100 a.u.. In figure A ($\delta$ vs $k$) the dashed line is a straight line starting at $\pi$ that is tangent to the ALDA curve. The solid lines correspond to  the accurate quantum chemical data from Ref.~\cite{S61}. In figure B we show the same data but now plotted against energy.}
%\end{figure}
 We report the scattering lengths we obtained from our phase shifts, or in the case of the full triplet ALDA calculation, from fitting a tangent to the curve, required to pass through $\pi$. In the triplet case, the value obtained from the fit agrees well with the reference value, as does the SPA result. Thus either method yields accurate results as $k\to0$.

Scattering from neutrals is very different from scattering from positive
ions.   In the former, the $N+1$-electron system has a {\em short}-ranged
potential, and so a finite cross-section,
but in the latter, the KS potential is {\em long}-ranged, i.e., it decays
as $-1/r$ for large $r$, and the cross-section diverges.  The phase-shift is
then defined relative to pure Coulomb scattering.  Our general approach
still applies, but Eq.~(\ref{e:phase}) must be modified.  If a potential
deviates from $-1/r$ only for $r<R_c$,
\ben
\tan(\delta_\alpha) = -F_l(k_\alpha R_c)/G_l (k_\alpha R_c)
\een
where $F_l$ and $G_l$ are the Coulomb
scattering solutions~\cite{F98}.  We will report results for positive ions
in future publications.

%\textcolor{blue}{\bf
%Anther point we make here is that we have only solved the TDDFT within the single-channel
%approximation, i.e., we ignore all matrix elements coupling different $l$ values.
%We believe that the simple formula, Eq.~(\ref{e:phase}), is
%exact only for potential scattering (despite being used more generally in the
%literature), and so only within a single-channel approximation
%for the many-electron case.  We have not yet attempted a correct treatment of
%the multi-channel case.
%THIS NEEDS TO BE CHANGED}

While the ALDA functional uses only input from the uniform electron gas, our results show that it gives accurate results for electron-scattering from a system that could not be further away from a homogeneous gas, the hydrogen atom. These results encourage us to continue work along these lines.
 We will calculate other $l$-values, different approximate ground-state potentials, different XC-kernels, other atoms, and ions, to gain experience in the reliability of TDDFT calculations. But we finish by considering some obstacles in applying our method to scattering from 
large molecules.  We first note that, by converting the problem to one of discrete
transitions, one needs only modify an existing electronic structure code by
placing a hard wall around it, rather than use a scattering code.  However, our
formula is only exact if the wavefunction has a node on the hard-wall surface,
which would only be true state-by-state for a non-spherical system.  Much better
is to use a large sphere, so that the formula is approximately true.
We must also address the multichannel case.  We intend applying our method
to electron scattering from Be$^+$ next, which is a well-studied scattering example~\cite{G81},
and for which the exact ground-state KS potential is known~\cite{UG93}.

We gratefully acknowledge support from NSF Grant No. CHE-0355405. MvF acknowledges The Netherlands Organization for Scientific Research (NWO) for support through a VENI grant.  We also acknowledge the exact H$^-$ KS potential from Cyrus Umrigar,  and useful discussions with Mike Morrison, Neepa Maitra, Chris Greene, Peter Lambropolus and Vladimir Mandelshtam.

%\bibliography{myrefers}

\end{document}